\documentclass[12pt,letterpaper]{article}
\usepackage[a4paper, total={7in, 10in}]{geometry}

\usepackage{graphicx}
\usepackage{helvet}
\usepackage{authblk}
\usepackage{hyperref}
\usepackage{amsmath} 
\usepackage{amssymb} 
\usepackage{orcidlink} 
\usepackage[super,comma,sort&compress]  
   {natbib}
\usepackage[right]{lineno} \linenumbers

\makeatletter
\renewcommand{\maketitle}{\bgroup\setlength{\parindent}{0pt}
\begin{flushleft}
  \textbf{\@title}
  
  \@author
\end{flushleft}\egroup}
\makeatother

\title{Safety challenges of AI in medicine in the era of large language models}
\date{}

\author[1,2,$\dagger$,*]{Xiaoye Wang}  
\author[3,$\dagger$]{Nicole Xi Zhang}   
\author[1]{Hongyu He}
\author[1]{Trang Nguyen}
\author[4]{Kun-Hsing Yu}
\author[4,5]{Hao Deng}
\author[6]{Cynthia Brandt}
\author[4,7]{Danielle S. Bitterman}
\author[8]{Ling Pan}
\author[1]{Ching-Yu Cheng}
\author[9]{James Zou}
\author[1,**]{Dianbo Liu}

\affil[1]{National University of Singapore, Singapore, Singapore}
\affil[2]{University of Cambridge, Cambridge, UK}
\affil[3]{Mila-Quebec AI Institute, Montreal, QC, Canada}
\affil[4]{Harvard Medical School, Boston, MA, USA}
\affil[5]{Massachusetts General Hospital, Boston, MA, USA}
\affil[6]{Yale University, New Haven, CT, USA}
\affil[7]{Brigham and Women's Hospital/Dana-Farber Cancer Institute, Boston, MA, USA}
\affil[8]{The Hong Kong University of Science and Technology, Hong Kong, China}
\affil[9]{Stanford University, Stanford, CA, USA}

\affil[$\dagger$]{These authors contributed equally}

\affil[*]{Correspondence: xw453@cam.ac.uk}  
\affil[**]{Correspondence: dianbo@nus.edu.sg; Lead contact}  

\begin{document}

\maketitle

\section*{SUMMARY}

Recent advancements in artificial intelligence (AI), particularly in large language models (LLMs), have unlocked significant potential to enhance the quality and efficiency of medical care. By introducing a novel way to interact with AI and data through natural language, LLMs offer new opportunities for medical practitioners, patients, and researchers. However, as AI and LLMs become more powerful and especially achieve superhuman performance in some medical tasks, public concerns over their safety have intensified. These concerns about AI safety have emerged as the most significant obstacles to the adoption of AI in medicine. In response, this review examines emerging risks in AI utilization during the LLM era. First, we explore LLM-specific safety challenges from functional and communication perspectives, addressing issues across data collection, model training, and real-world application. We then consider inherent safety problems shared by all AI systems, along with additional complications introduced by LLMs. Last, we discussed how safety issues of using AI in clinical practice and healthcare system operation would undermine trust among patient, clinicians and the public, and how to build confidence in these systems. By emphasizing the development of safe AI, we believe these technologies can be more rapidly and reliably integrated into everyday medical practice to benefit both patients and clinicians.
\section*{KEYWORDS}


Artificial Intelligence, AI for Healthcare, Safety, Large Language Models

\section*{INTRODUCTION}

Recent advances in artificial intelligence (AI) have created significant opportunities to enhance the efficiency and quality of healthcare. In 2023 alone, nearly 700 AI-enabled devices were authorized by the U.S. Food and Drug Administration, spanning various medical fields including radiology, ophthalmology, and hematology \cite{fdaAIML}. Despite rapid technological progress, AI is still not widely used in real-world healthcare. The main barrier is the safety concerns shared by patients, clinicians, and the public. A recent U.S. survey revealed that 60\% of the population feels uncomfortable with healthcare providers relying on AI \cite{pewAI2023}.

Large language models (LLMs) are advanced AI systems designed to understand and generate text with near-human fluency. Over the past two years (2022–2024), several LLMs—such as ChatGPT, Gemini, and Claude—have emerged and gradually been adopted in various medical fields. Recently, they have demonstrated superhuman performance in tasks like clinical text summarization and mental health disorder identification \cite{van2023clinical}. Despite demonstrating excellent capabilities and promising prospects in various medical tasks, the distrust towards medical AI models further intensify when people realize these machine intelligence can, in some cases, surpass human experts-an unprecedented development in human history.

Thus, the key to fostering the widespread adoption of AI in everyday healthcare practice is building confidence and trust among patients, clinicians, and the broader public. In this article, we systematically reviews safety issues of AI in medicine in the era of LLMs and discuss how possibly trust on AI can be won in healthcare settings. First, we examined the AI safety issues that are unique with LLMs. Next, we reviewed safety challenges common to both traditional AI models and LLMs, highlighting the distinctions in the context of the LLM era. Finally, we discussed how these safety concerns can undermine trust in clinical practice and the operation of healthcare systems.


Certain safety issues are significantly more common in LLMs compared to other AI methods due to their interaction with humans through natural language and their construction using techniques such as self-supervised training, fine-tuning, and reinforcement learning with human feedback. In Section 1\ref{section1}, we categorize LLM-specific safety issues into two main areas. The first category relates to functionality-based safety issues, focusing on LLMs' data, training stage, model architecture and capability, fine-tuning, and inference. The second category encompasses communication-related safety problems, primarily concerning human-LLM interactions. These safety issues comprehensively summarize the potential problems that LLMs may encounter in the medical domain throughout the entire process - from organizing training data and designing training models to the final model deployment and utilization.

For general safety issues shared by common AI models and LLMs, We identify two key areas where safety concerns arise in Section 2\ref{section3}: reliability and alignment. The complexity of AI reliability encompasses issues such as data harmonization, consistent performance, calibration, generalization, bias and fairness, and domain adaptation \cite{rajkomar2019machine,yu2019framing}. Regarding AI alignment, in this article, this refers to ensuring AI adheres to human-defined objectives and principles, addressing concerns like objective mis-specification and hacking reward set by patients or clinicians, as well as AI-human interaction.

Building on our analysis of the aforementioned safety issues from a technical perspective, we further explored how these issues undermine trust from a medical standpoint, particularly for healthcare professionals and researchers using AI models and LLMs in tasks such as assisted diagnosis in Section 3\ref{section3}. This analysis covers safety concerns in clinical practice, healthcare system operations, and societal impacts, truly examining the influences and risks these emerging technologies bring to the healthcare system from the users' perspective.



\section* {Unique AI Safety Challenges of Large Language Models in Medicine} \label{section1}

\begin{figure}[h]
    \centering
    \includegraphics[width=0.8\linewidth]{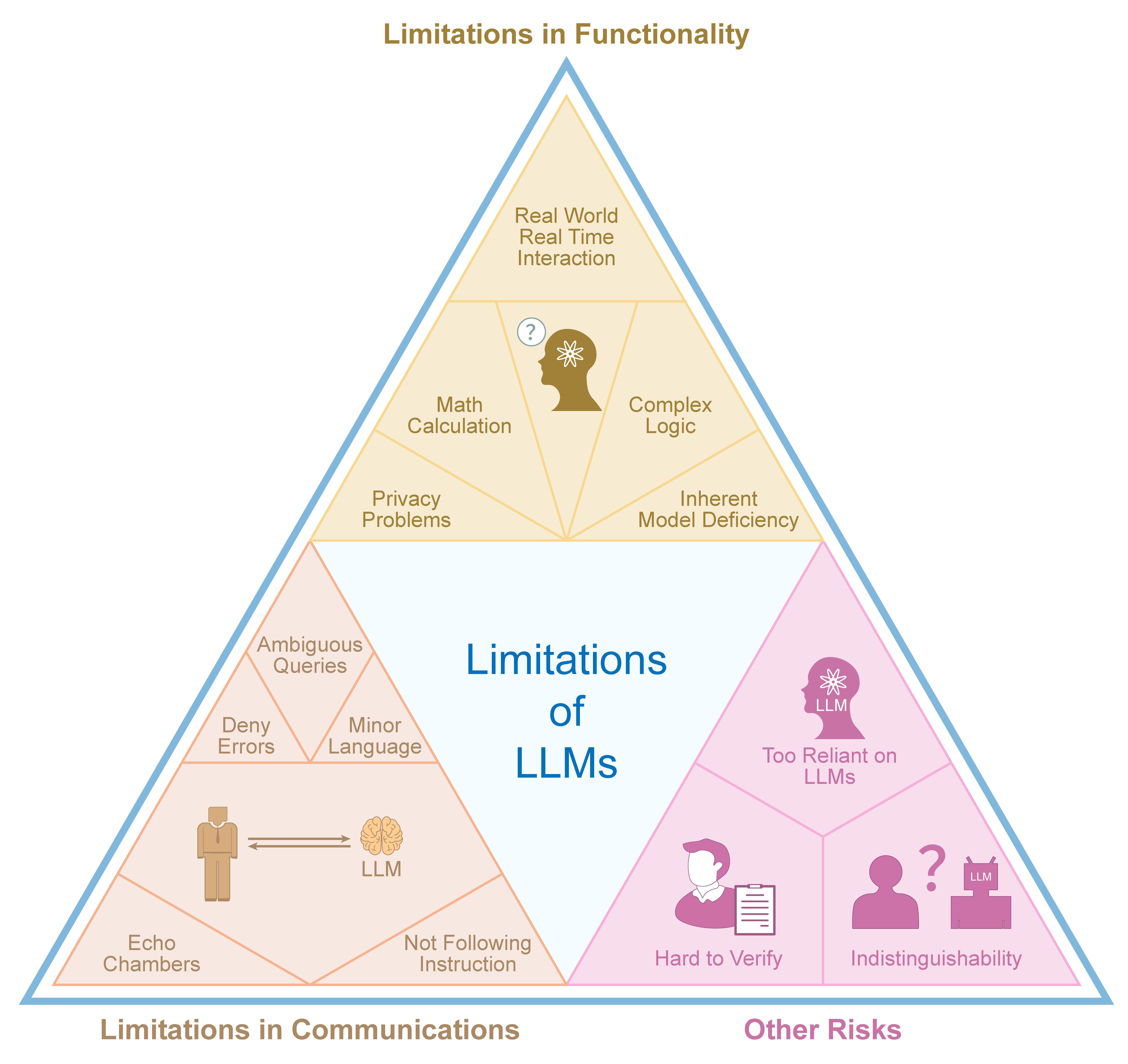}
    \caption{AI safety challenges in medicine related to  large language models.}
    \label{fig:llm_issues}
\end{figure}

LLMs like GPT-4, GPT-o1, Claude 3.5, and Llama 3 have a wide range of applications in medicine. They have been adapted for clinical documentation and reporting, medical information retrieval, disease prediction and diagnosis, and drug discovery and development and many other tasks \cite{singhal2023large, yang2022large, agrawal2022large, chakraborty2023artificial, zhang2024generalist, abramson2024accurate}. In recent years, a combination of a shortage of well-trained physicians and the increased complexity in the medical field constitutes a significant challenge for the timely and accurate delivery of healthcare. LLMs have been tested to diagnose complex clinical cases and show better diagnosis than human annotators \cite{eriksen2023use}. Therefore, LLMs, together with other AI models, could possibly be the key solutions to improve efficiency in healthcare. However, as a new interface between humans, data and computers, LLMs have many unique safety issues that are rarely found in other types of AI models. Over the past year leading up to the publication of this manuscript, we have witnessed a surge in research dedicated to enhancing the safety of LLMs from various angles.  In this section, we aim to discuss different problems of LLMs that may lead to potential medical risks when applied in healthcare, as summarized in Figure~\ref{fig:llm_issues}.

\subsection*{Safety challenges caused by functionality of LLMs}

It is pointed out that when using LLMs such as GPT-4 for medical purpose, two major safety problems are \cite{goldberg2023patient}: ``Do not trust it. This type of AI makes things up when it does not know an answer. Never ever act on what it tells you without checking" and ``You should know, though, that if you paste medical information into an online AI program, it loses our privacy protections", The first point is a problem referred to as ``hallucination" in the AI community. The second point is the privacy issue of sensitive medical data.



In the field of AI, a \textit{hallucination} or artificial hallucination \cite{alkaissi2023artificial} is a confident response by an AI that does not seem to be justified by its training data or real-world facts. In order to answer the request of the user, LLMs may make up responses that may not be true. Hallucination about diagnoses, treatment and patients outcomes in texts generated by LLMs will confuse and mislead decisions by both clinicians and patients. 
\textit{Math calculation ability} of LLMs has been recently reported to be an issues in medicine. According to Barker et al \cite{barker2002medication}, one in five medication doses during hospital stages are given with errors, and more than 7 million patients per year in the US are affected by such errors \cite{da2016alarming}. There, AI and LLMs hold huge opportunities to improve these issues. Recent work shows that LLMs, such as GPT-o1, is able to conduct calculations or code calculator app, indicating their potential to revolutionize the way healthcare professionals approach medication dosing and reduce the likelihood of errors. However, the integration of these advanced technologies in clinical settings is not without its challenges. Prof. Kohane's observation about GPT-4's inaccuracies in calculating the Pearson correlation for a patient's salt intake and systolic blood pressure serves as a critical reminder of the limitations and pitfalls associated with relying solely on AI for healthcare decisions \cite{lee2023ai}. Such discrepancies highlight the importance of incorporating an additional layer of verification, particularly for complex mathematical tasks that have direct implications for patient care. 

Concerns were also raised about current LLMs' difficulty to \textit{handle complex logic}. According to  Dziri et al \cite{dziri2023faith}, through multi-digit multiplication, logic grid puzzles, and a classic dynamic programming problem, they show empirically that LLMs cannot handle compositional tasks that require breaking down  problems into compositional sub-steps and synthesizing these steps into a precise answer. In fact, transformer models solve compositional tasks by reducing multi-step compositional reasoning into linearized subgraph matching without necessarily developing systematic problem-solving skills. This will lead to unreliable results for compositional tasks in medical settings such as treatment design that requires multiple steps of logic reasoning based on different observations. 
Everyday medical operation requires strong \textit{common-sense reasoning} ability of the AI models. As reported by Lee et al. \cite{lee2023ai}, GPT-4 is able to have amazing performance in reasoning how to transfer different patients to different destinations and infer thoughts of the nurse who is arranging the transfer. However, some studies show limitations and instability of performance of LLMs in various common sense reasoning tasks such as in physical relation among household items\cite{agrawal2023physical}. If we do not have a way to estimate LLMs' limit, and when they can be trusted, it will be very difficult to develop robust LLM based tools to handle important medical use cases such as arrangement of medical devices.


\textit{Interaction with private patient's Information} always requires the highest level of caution. As stated by ChatGPT itself, the models don't explicitly memorize input data. However, there are still concerns about potential data leakage or the possibility of extracting specifics about the training data. In the most recently deployed LLMs, local users do not have access to the model but need to access the model by providing queries and contextual data to an online portal. In healthcare, most data belongs to individual patients and is protected by law \cite{kim2023propile}.

Another challenge in LLMs is \textit{physical world understanding}. When inputting the question ``Can you describe how you perceive or experience the physical world?" to ChatGPT, we got the following answer: ``I don't have sensory experiences. While I can provide information about the physical world based on my training data, I don't `experience' or `understand' it as humans do." Even visual-large language models such as ChatGPT-V became available \cite{wu2023visual}, we still lack an efficient way to enable LLMs to interact with the physical world, such as asking questions like ``Which of the three ventilators in the next room should I use for the patient seen by Jone yesterday morning in his clinic?", which limits the usage of LLMs in many healthcare related activities.

\textit{Fact-checking in real-time} is also a problem of using LLMs in medicine. The users can not verify the real-time accuracy of statements or determine the truthfulness of recent claims, such as generated clinical summaries. The user sometimes can rely on the training data for factual information but not all the time. 
As many LLMs are trained offline, the \textit{lack of up-to-date information} prevents the use of LLMs in healthcare operations which require frequent updates of entirely new information. For instance, the information used to train models like ChatGPT is typically a few months old. In the healthcare sector, where current information is crucial, the slow pace at which models are updated by their developers poses a significant limitation. Therefore, hospital users often need to tune their own version of LLMs either using service from large model providers or based on open source models \cite{li2023chatdoctor}. 

\textit{Context length} is another challenge in some LLMs. Even though many LLMs and related transformer architectures have shown significant advancements in understanding and generating complex language structures, there is a notable limitation in their context length capabilities, even with recent increases in the newer version such as GPT-4o and GPT-o1 \cite{masumura2021hierarchical,pope2023efficiently}. This constraint becomes especially apparent when comparing the memory of LLMs to humans, particularly across diverse conversations and documents. 
The limitation of \textit{accessible data types} poses a another challenge for clinicians, patients, and researchers looking to leverage LLMs in medical practice or research. Currently, these models predominantly excel with textual data and images, which they can analyze and interpret with a high degree of accuracy. However, when it comes to other crucial types of data in the medical field, such as EEG waves, protein 3D structures, or genomic sequences, ordinary LLMs face limitations. These data types embody complex patterns that require not only advanced processing to convert them into a format understandable by LLMs but also necessitate the development of specialized algorithms capable of capturing the intricacies inherent to such information. Healthcare data is inherently multimodal and longitudinal, with only a fraction of data in structured forms. Rather than textual and visual data, most large generative models, including GPT-4V and Gemini, are not able to process a large number of different data modalities. Even a model is adjusted for different data modality, researchers still lack of a reliable set of benchmark to systematically evaluate model performance in medical settings as evaluating generative AI models is challenging. The vast amount of routinely collected real-world clinical data, such as electronic medical records or remote sensing data, represents a significant opportunity for the growth of multimodal AI models. Yet, the AI for medicine community still searches for effective methods to harness these multimodal real-world datasets.

\begin{table*}[htbp]
\caption{Limitations of LLMs in communications.}\label{table4}
    \centering
    \scalebox{0.92}{
        \begin{tabular}{p{3.5cm}|p{13.5cm}}
            \hline 
            \hline 
            
            {Potential for Echo Chambers} & 
            If users primarily interact with models that ``agree" with them or provide information that aligns with their beliefs, it could further entrench existing viewpoints and reduce exposure to diverse perspectives. \\ \hline
            
            {Not Following Instruction} &
            At times, LLMs may choose not to follow instructions by the user closely. It is reported by Lee et al. that when an early version of GPT-4 was instructed to generate a document with certain patterns, it generated a different document due to a preference \cite {lee2023ai}. This will result in producing unwanted information, which leads to misleading medical decisions. \\ \hline
            
            {Denial on Errors} &
            In certain cases, even if the user points out the mistake by the LLMs, the AI may deny its mistake and insist on the wrong results. One such example the wrong Pearson correlation between salt intake and blood pressure until Prof. Kohane provides a math equation for Pearson correlation\cite {lee2023ai}. \\ \hline
            
            {Sensitivity to Ambiguous Queries} &
            If a question is ambiguous or lacks context, LLMs might infer the user's intention incorrectly and provide an answer that doesn't align with what the user had in mind\cite{zhang2024clamber}. \\ \hline
            
            {Documents in minor languages} &
            The amount of information available to train LLMs varies among languages; LLMs may perform better in one language than another which may lead to unfair performance and quality of healthcare for clinicians and patients using different languages \cite{ramesh2023fairness}.
            \\\hline	
    \end{tabular}}
\end{table*}



Moreover, unlike many other fields, it is extremely challenging to gather a huge amount of relevant \textit{data with high quality} for specific practices in healthcare, such as dentistry.  Despite rigorous efforts to sanitize and filter the vast amount of training data, it takes a significant amount of efforts, if not impossible, to eliminate all harmful and inappropriate content, which may inadvertently propagate through the responses generated by LLMs. Inherently, these LLMs operate as sophisticated pattern-matching machines without a genuine understanding of the data they are trained on, which occasionally leads to nonsensical or inappropriate responses with erroneous domain knowledge \cite{huang2023chatgpt,mcnichols2023algebra}. 

\subsection*{Safety challenges caused by communications between human and LLMs}

Beyond the aforementioned safety issues in LLMs' data, training, and inference, how we use the models and establish effective communications with LLMs also pose potential risks for their safe application in the medical domain.
When interacting with LLMs, we may not always receive the outcome we want due to these \textit{limitations in communication} between human users and the models.
These limitations include and are not limited to LLMs acting as a echo chambers and not following instruction (see more in Table \ref{table4}).

In addition to the risks mentioned in previous sections, many other potential risks will emerge as LLMs become increasingly powerful and widely implemented in medical practices. Some examples include over-reliance on LLMs, difficulty in verifying the information generated by the models, and the challenge of distinguishing between generated and human-written information.

\begin{figure}[h]
    \centering
    \includegraphics[width=1.0\linewidth]{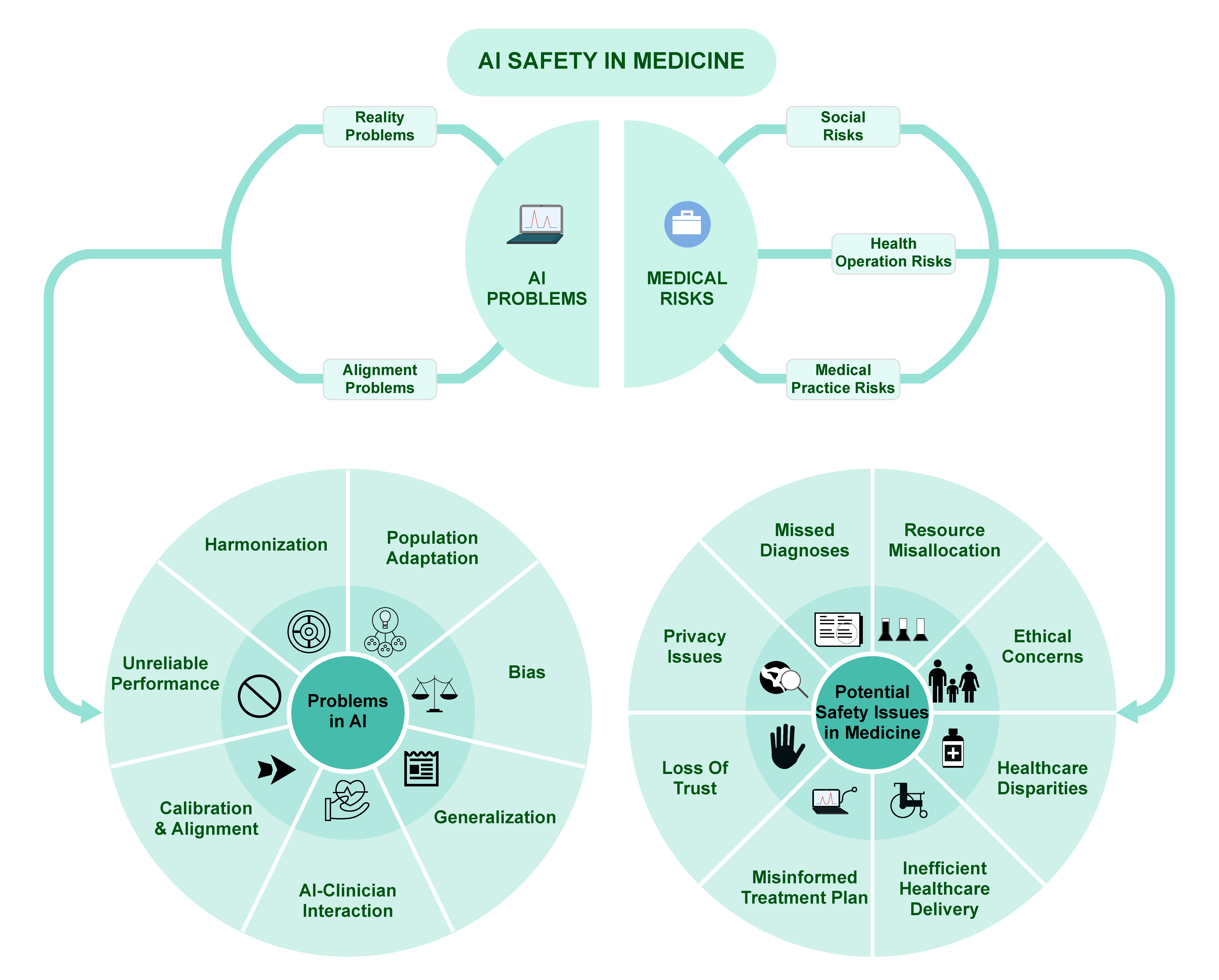}
    \caption{General safety issues in medicine shared by common AI models and LLMs. These are inherent problems of most AI models related to real-world healthcare.}
    \label{fig:Structure}
\end{figure}

\section*{General Safety Challenges of AI Models in Medicine in the Era of LLMs } \label{section2}

Beyond the safety concerns specific to LLMs discussed in the previous section, many challenges shared by all AI methods also raise safety concerns in medicine. In this section, we delve into the reliability and alignment challenges inherent to general AI models that have significant implications for medicine, while also exploring emerging issues unique to the era of LLMs.


\subsection*{Reliability problems of artificial intelligence in medicine}
In the medical practice, building general AI models and LLMs that can be relied by patients, clinicians, and the general public is important, given the high stakes involved in healthcare decisions and outcomes. However, obtaining a reliable model, which means the established model should be invariant to potential disturbances and always provide the correct and accurate guidance towards medical requirements regarding different inputs, is challenging due to various issues. In this section, we discuss five challenges in medical settings including data harmonization, model calibration, momdel generalization, biases, and the difficulty in adapting to a new patient population.

{}{Data harmonization challenge} originates from the heterogeneity of data used to train AI models. The exponential growth in data volume is accompanied by increased diversity, presenting significant challenges for data management and analytics. The task of sifting through such heterogeneous data is particularly complex, affecting the processes of data visualization and prediction, and consequently the analytical outcomes. Data harmonization is a crucial process aimed at standardizing the disparate forms of data for uniform interpretation especially in medicine \cite{kumar2021data}. In the domain of AI for medicine, models trained on limited datasets will lack generalizability. The current trend gravitates towards the use of extensive datasets amassed from multiple healthcare institutions, making the harmonization of multi-institutional data indispensable, especially for developing LLMs.
This harmonization is often the preliminary step in data science projects. Inadequately done, it can introduce errors that lead to grave medical repercussions. Regardless of scale, the errors invariably alter feature values and distributions, with some features less tolerant to such perturbations. As a result, combining these features for statistical analysis and model building could yield unreliable outcomes, either by obscuring true correlations or by falsely indicating the presence of correlative relationships \cite{papadimitroulas2021artificial}.

The harmonization of medical data for AI applications encounters several substantial challenges including: 
(1) \textit{Diverse Data Sources}: Medical data comes from numerous sources, such as Electronic Health Records (EHR), medical imaging, genomic data, and wearable device data. Their unique structures, formats, and representations make it difficult to integrate and analyze collectively \cite{tiwari2020assessment,selim2021cross,lee2017medical}.
(2) \textit{Inconsistency and Variability}: Different healthcare providers may use different terminologies, measurement units, and encoding schemes for the same data type, thus creating discrepancies and inconsistencies in the datasets. Most data have been shown to be highly sensitive to inter-institutional variations in several factors, including variability in device manufacturer, generation and models, acquisition protocols, and reconstruction settings \cite{papadimitroulas2021artificial}.
(3) \textit{High Dimensionality and Complexity}: Medical data is typically high-dimensional, involving numerous features and variables. The complexity increases when different data types need to be combined for analysis, requiring sophisticated methods to identify patterns and make predictions \cite{glocker2019machine}. 
(4) \textit{Lack of Standardization}: Even though there exit certain standards for medical devices and industry such as ISO, the lack of widely recognized and used global standards in medical documentation and representation of data, introduces significant variability in data quality and structure, posing difficulties in data integration and subsequent harmonization efforts \cite{saripalle2019using,liu2019high}.
(5) \textit{Privacy and Security Concerns}: Medical data is inherently sensitive, and its management is bound by stringent privacy and security regulations. These regulations, while needed to protect patient privacy, often limit the extent of data sharing and collaboration, thereby constraining the scope of harmonization processes \cite{kaissis2020secure,price2019privacy,cui2021fearh}.
(6) \textit{Bias and Representativeness}: Discrepancies in data collection methods and patients' access to healthcare systems can result in biased or unrepresentative datasets. Such datasets skew the development of AI models, adversely affecting their generalizability and reliability \cite{vokinger2021mitigating,gianfrancesco2018potential}. These complications collectively underscore the complexity of creating cohesive and comprehensive medical datasets suitable for developing reliable and effective AI applications in healthcare. In the era of LLMs, errors in the data harmonization process can have significantly more severe consequences. The construction of most LLMs involves a pretraining or self-supervised learning phase, during which models derive patterns from vast amounts of unlabelled data. More recently, many AI models have been developed—either entirely or partially—using synthetic data generated by LLMs. These practices exacerbate the impact of imperfections in the medical data harmonization process, amplifying their potential to propagate errors in healthcare.

{}{Model calibration} is another major challenge in AI for medicine. Model calibration in machine learning is a critical process that ensures the predicted probabilities of an outcome align closely with the actual occurrence rates of that outcome. For instance, if a calibrated model predicts an event with a probability of 80\%, then in a large patient population, that event should manifest approximately 80\% of the time. Consider the following example to illustrate this concept: a model analyzes a group of patients' medical history, genetics, and lifestyle factors. For Patient A, the model predicts a 70\% chance of developing a certain disease within the next year. For Patient B, the prediction is a 20\% chance. For a well-calibrated model, the predicted risk probability for a large cohort of patients should be similar to the empirical observation, therefore approximately 70\% of patients similar to patient A will develop the disease. 
The process of calibration ensures that the predictions made by AI models in medical contexts are reflective of real-world probabilities. The primary challenge in model calibration in the medical field is the proper interpretation and application of confidence levels associated with AI predictions. 
First, there is a \textit{lack of standardized calibration benchmarks}, as no universally accepted guidelines exist for calibrating AI models in medicine. This absence leads to varying levels of model reliability and complicates performance assessment across healthcare institutions. 
Second, \textit{misinterpretation of confidence levels} is common in medical settings. Clinicians may over-rely on AI predictions with high confidence, overshadowing clinical judgment, or underutilize valuable predictions due to a lack of trust or understanding, especially when models provide moderate but clinically significant predictions. 
Lastly, the \textit{evolving nature of medical knowledge and practices}, along with shifting disease prevalence, makes maintaining AI model calibration challenging, necessitating continuous updates and recalibration to ensure accuracy with new treatments and diagnostic techniques.

In addition to these shared calibration challenges, LLMs also present unique calibration safety concerns. As highlighted by Xie et al. \cite{xie2024survey}, accurately assessing the reliability of LLM outputs remains a persistent challenge due to their inherent overconfidence and susceptibility to producing hallucinations. For widely-used black-box LLMs like GPT, Claude and Gemini, which only permit API-level interactions, the lack of access to internal model parameters exacerbates the calibration difficulty. Moreover, standard methods like logit-based calibration \cite{bi2024decoding} or instruction tuning \cite{kapoor2024large}, effective in white-box machine learning and LLM settings where the checkpoints and model architectures are accessible, are often impractical for black-box models. Furthermore, LLM-specific calibration issues arise from the nuanced nature of language tasks. These models may exhibit overconfidence in ambiguous or low-context scenarios, making it difficult to align confidence scores with actual response correctness. Strategies like post-processing confidence estimation \cite{zhang2024luq, li2024confidence}, employing proxy models \cite{ulmer2024calibrating, shen2024thermometer}, and integrating multi-model collaboration frameworks \cite{feng2024don, zhang2023sac} have been proposed to address these challenges, but each comes with its limitations, particularly in terms of scalability and adaptability to evolving knowledge bases as mentioned.

\begin{figure}
    \centering
    \includegraphics[width=0.6\paperwidth]{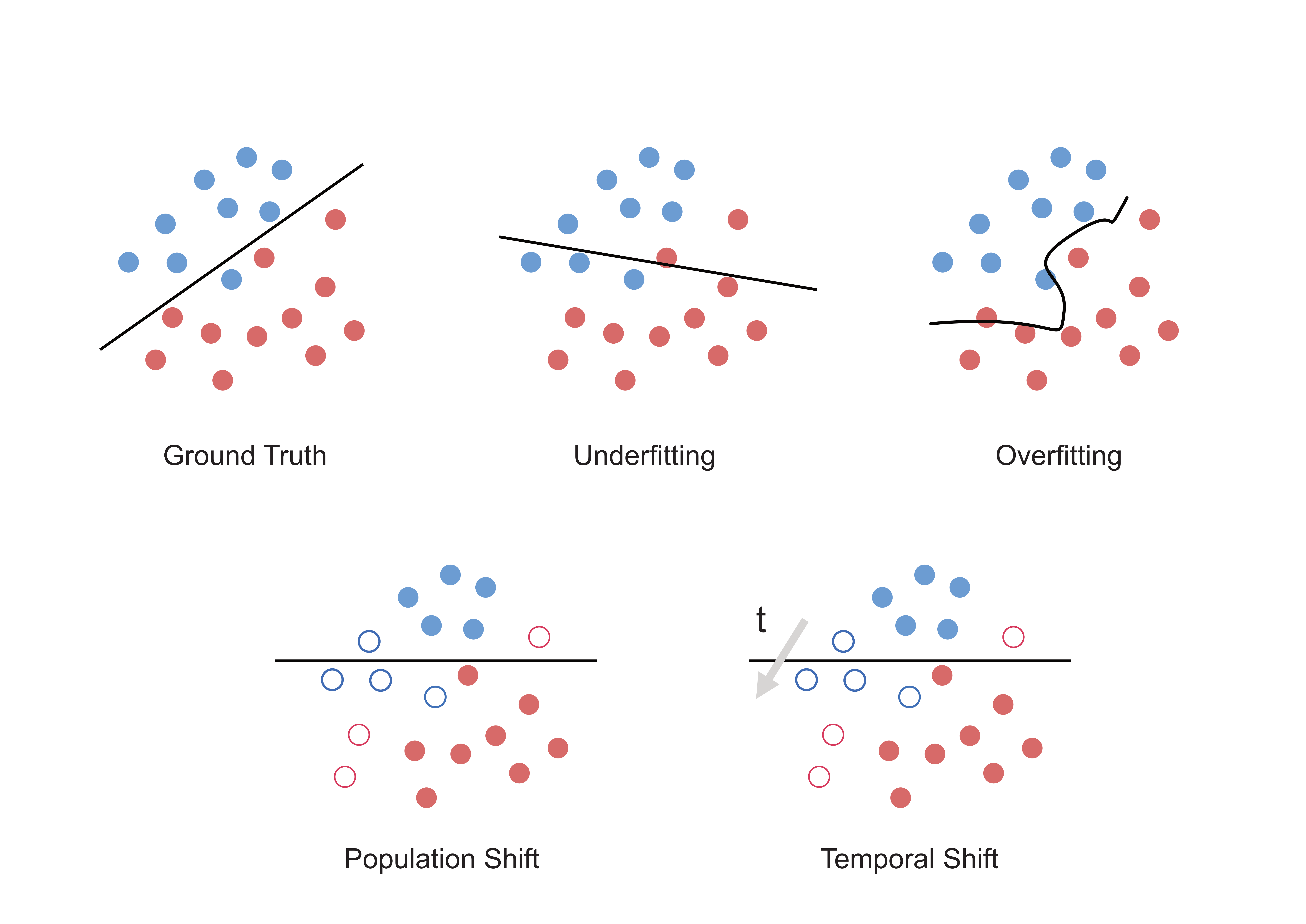}
    \caption{Generalization issue visualization of AI in medicine. This figure illustrates how AI models perform differently on training data versus new, unseen data. The correct model is shown as the ground truth in the top left. Underfitting occurs when the model is too simple to capture patterns. Overfitting happens when a model performs well on training data but poorly on new data. Population shift refers to a model's performance decline when applied to patient groups with different data distributions. Temporal shift occurs when patient data distributions change over time.}
    \label{fig:generalization}
\end{figure}

Generalization issues can be another obstacle for safely adopting general AI models and LLMs in real-world medical settings. In machine learning, {}{generalization} is a model’s ability to perform well with unseen data after training on known data (see Figure \ref{fig:generalization}).
The concept of generalization is borrowed from psychology, where learned knowledge can be transferred to novel scenarios \cite{Shepard1987}. 
For instance, a child quickly knows a cheetah is a cat without seeing many cats. Machines lack human intuition but can learn from vast amounts of data. It’s crucial to have comprehensive data about a subject to make model that can generalize.
Strict regulations on patient health data prevent many ML-driven healthcare studies from being evaluated on external patient groups, resulting in discrepancies in the model's performance across different sites. Most data sources are confined to specific institutions or regions. A study discovered that only about 23\% of healthcare-related ML articles utilized multiple datasets \cite{yang2022machine}. In some instances, the error rate of a deep learning model for retinal image analysis was 5.5\% on the training dataset but increased to 46.6\% when tested with images from another vendor. The problem of generalizability has become one of the major roadblocks to deploying deep learning models into clinical practices \cite{de2018clinically,zhang2020generalizing}. Several types of generalization issues commonly arise in the application of AI in medicine: (1) \textit{Overfitting}: This occurs when a model memorizes the specifics of the training data, including noise and outliers. Consequently, while it performs well on the training dataset, its effectiveness diminishes significantly on new, unseen data. (2) \textit{Underfitting}: The converse of overfitting, underfitting happens when a model is too simplistic, failing to capture the underlying complexities and patterns in the data. This leads to poor performance on both the training and test datasets. (3) \textit{Population Shift}: If the training dataset lacks diversity and does not sufficiently represent various patient populations, the AI model's ability to generalize effectively to underrepresented groups is compromised. (4) \textit{Temporal Changes}: The dynamic nature of medical practices and patient demographics means that models trained on historical data might not perform effectively in contemporary or future scenarios.

While LLMs have alleviated some of generalization issues in common AI models such as overfitting and underfitting through scaling, they still face unique safety challenges related to data shifts, which include both population shifts and temporal shifts \cite{datashift2024}. 
In terms of population shifts, even though LLMs have seen a huge amount of training data encompassing a wide range of distributions, considering that most datasets used to train medical LLMs are limited to specific countries or regions rather than encompassing global medical data, applying LLMs trained on data from region A to region B significantly reduces their accuracy and generalization performance. 
These demographic shifts or under-representation of certain patient populations in the training data can compromise the model’s reliability across diverse groups, limiting the applicability of LLMs in real-world medical scenarios.
For example, a multimodal disease prediction LLM trained on color fundus images and EHR data from the European countries would achieve only slightly better accuracy than random guessing when applied to Asian populations. The innate racial differences and acquired lifestyle differences between regions introduce significant data shifts for LLMs, posing substantial risks to their safe application.
In addition to the above, temporal changes further compound this problem by introducing dynamic shifts in the relevance and accuracy of past data \cite{mousavi2024dyknow, price2024future, zhu2024your}.
As highlighted by Zhu et al.\cite{zhu2024your}, LLMs often exhibit nostalgia bias (over-reliance on historical data) and neophilia bias (overemphasis on recent trends), which hinder their ability to generalize effectively across temporal contexts. These limitations are further exacerbated by the static nature of pretraining datasets, making it difficult for LLMs to adapt to evolving knowledge and trends.
In medicine, the introduction of new terminologies, guidelines, and diagnostic techniques can quickly render an LLM’s training data outdated, leading to decreased prediction accuracy and inconsistent outputs. For instance, models trained on historical medical data might struggle to adapt to emerging diseases or updated treatment protocols, potentially resulting in unsafe or misleading predictions.

The problem of \textit{bias} in AI model performance in medicine is an important concern as it impacts the reliability and fairness of medical decisions, potentially leading to suboptimal care and health disparities among different population subgroups \cite{rajkomar2018ensuring,fletcher2021addressing, Obermeyer2019}. The development and deployment of AI in medicine faces several bias-related challenges: 
(1) \textit{Label Bias}: The labels in training data might reflect existing biases in clinical practice, mirroring the unequal attention and care provided to different demographic groups, thereby influencing AI outcomes. 
(2) \textit{Algorithmic Bias}: The choice of algorithms, features, and model parameters can introduce and amplify biases. This affects model predictions and outcomes, potentially leading to inequitable healthcare delivery. 
(3) \textit{Evaluation Bias}: Evaluation metrics and benchmarks might be biased towards certain groups or conditions. This can mask the true performance of AI models across diverse populations. 
(4) \textit{Socioeconomic and Cultural Bias}: Differences in access to healthcare resources, healthcare-seeking behaviors, and disease prevalence across socioeconomic and cultural groups can lead to bias in construction of AI models. 
(5) \textit{Historical Bias}: Historical inequalities and prejudices in medical research and clinical practice can be perpetuated and amplified through biased AI models, necessitating a conscientious approach to model training and validation.

While general AI models face various bias-related challenges, LLMs introduce additional bias safety concerns that warrant careful consideration in medical applications \cite{schmidgall2024addressing, peng2024securing, poulain2024bias}. Firstly, the expansive training datasets of LLMs may embed gender, race, societal, cultural, institutional and political biases, which can inadvertently perpetuate stereotypes or inequities in medical contexts \cite{schmidgall2024addressing, peng2024securing, zack2024assessing, omiye2023large, seshadri2023bias, clarke2023protests, garcia2023uncurated}. For instance, LLMs may demonstrate geographic bias, favoring regions with greater representation in the training data, potentially leading to inaccurate or discriminatory recommendations for underrepresented areas. 
Furthermore, as illustated by Schmidgall et al. \cite{schmidgall2024addressing}, cognitive biases, such as self-diagnosis bias, confirmation bias, and recency bias, significantly influence the performance of LLMs in medical contexts. These biases can compromise diagnostic accuracy and decision-making, especially when models are exposed to prompts that emulate real-world scenarios involving clinical biases. The development of benchmarks like BiasMedQA emphasizes the critical need for tailored mitigation strategies for the safety of medical LLMs.
In addition, LLMs exhibit implicit and source biases due to the statistical patterns inherent in their training data, which can disproportionately favor familiar or frequent inputs while underrepresenting less common but critical medical cases. Such biases can compromise the reliability of LLMs in addressing rare diseases or conditions prevalent in specific populations. Biases introduced during fine-tuning or safety interventions can also lead to unintended trade-offs, such as diminished contextual accuracy or overgeneralization in nuanced scenarios.

Fine-tuning AI models to local data in medicine is essential to adapting models to the specific characteristics of new patient populations and healthcare settings. However, it poses several challenges that impact the effectiveness and generalizability of models \cite{laparra2020rethinking,zhang2020collaborative}. Apart from the  aforementioned data related challenges including \textit{Data Scarcity}, \textit{Data Quality and Variability}, \textit{Heterogeneity}, and \textit{Data Shifting}, \textit{Computational Resources} may also lead to safety issues, given that the limitation of computational resources in local environments can be a significant barrier to the fine-tuning of complex models. Moreover, \textit{Domain Expertise} can be a problem as well, as the absence of local expertise in fields like AI and medicine can pose challenges to the adaptation and validation of models.

In addition to these, \textit{Regulatory and Ethical Issues} also pose great challenges for general AI models when applied to a new patient population, and the emergence of LLMs with their broad range of applications further amplifies the severity of this issue\cite{ong2024ethical, mesko2023imperative, freyer2024future, haltaufderheide2024ethics}.
Firstly, data privacy and security concerns are paramount, as LLMs often rely on extensive datasets that may include sensitive patient information. Without rigorous safeguards, there is a risk of unauthorized data access, re-identification of anonymized data, or breaches during model deployment to new populations. The lack of transparency around the provenance of training data further complicates compliance with intellectual property and regulatory standards. Questions about whether datasets were appropriately licensed or whether their use respects patient consent add layers of complexity.
Additionally, the high variability and context-sensitivity of LLM outputs create challenges in meeting regulatory requirements for medical applications. Issues like hallucinations can have severe implications in clinical settings, such as misdiagnoses or inappropriate treatment recommendations. Moreover, existing regulations, such as the EU Artificial Intelligence Act \cite{down2021proposal}, may not fully address the unique capabilities and risks of LLMs, necessitating the development of more tailored frameworks.


\subsection *{Alignment problems of artificial intelligence in medicine}
Besides performing different tasks reliably and robustly, it is important to ensure the AI models' and LLMs' behaviors reflect human users' goals and values. In this section, we discuss the challenges of alignment problems for AI in medicine, with a focus on technical and normative alignment as well as AI-clinician Interaction.

The {}{alignment problem} in AI refers to the challenge of ensuring that artificial intelligence systems reliably behave in ways that are in accordance with human values, intentions, and expectations. This problem arises from the inherent difficulty in specifying complex human values and objectives precisely and completely in a manner that AI systems can understand and implement. The goal of AI value alignment is to ensure that AI as well as LLMs is properly aligned with human values \cite{noel2020human}.

The challenge of alignment has two parts. The first is technical, focusing on the formal incorporation of values or principles into AI to ensure their reliable and appropriate actions. Real-world instances of AI misalignment have been observed, such as chatbots promoting abusive content when interacting freely online \cite{wolf2017we}. Moreover, this challenge escalates with more advanced AIs, raising issues like `reward-hacking'—where agents find unanticipated methods to fulfill objectives, diverging from intended outcomes—and the difficulty of evaluating AIs whose cognitive abilities might significantly surpass human capabilities \cite{christiano2016prosaic}.

The second aspect of the value alignment problem is normative, raising the question of which values or principles should be encoded into AI. This can be viewed through two lenses: minimalist and maximalist conceptions of value alignment. The minimalist approach seeks to bind AI to a viable framework of human values to prevent unsafe outcomes, while the maximalist perspective aims for alignment with the most accurate or optimal human values on a broader societal or global scale. Although the minimalist perspective rightly notes that optimizing for nearly any single metric could yield adverse outcomes for humans, advancing beyond this approach might be necessary for achieving full alignment with AI. This is because AI systems could be safe and reliable but still fall short of optimal or truly desirable outcomes \cite{noel2020human}. Here we conclude the alignment challenges as in Table \ref{table2}.

\begin{table*}[htbp]
\caption{Alignment issues for AI in medicine.}\label{table2}
    \centering
    \scalebox{0.92}{
        \begin{tabular}{p{3.5cm}|p{13.5cm}}
            \hline 
            \hline 
            
            {Objective Mis-specification}& AI systems are designed to optimize specified objectives, yet these objectives may not fully capture intricate human values and goals. \\ \hline

            {Reward Hacking}& AI models risk discovering unintended shortcuts or exploits to maximize their reward function, diverging from the desired task.  \\\hline

            {Scalability and Complexity}& With increasing complexity and capability, ensuring that AI models remain aligned with human values and controlling their behavior become more challenging. \\\hline
            
            {Ethical and Value Alignment}& Balancing a wide array of human values, ethical considerations, and societal norms within AI systems represents a complex and often ambiguous endeavor. \\\hline

            {Long-term and Societal Impacts}&  Ensuring that AI systems align with broader, long-term societal objectives and mitigating their potential negative impacts on society and the environment pose multifaceted challenges.
            \\\hline	
    \end{tabular}}
\end{table*}

While general AI models face alignment challenges, LLMs encounter additional safety concerns due to their expansive capabilities and widespread applications\cite{han2024medsafetybench, liu2023trustworthy, anwar2024foundational}. 
One significant challenge lies in fine-tuning and adapting LLMs to domain-specific safety standards, such as those outlined by the American Medical Association’s Codes and Principles of Medical Ethics \cite{AMAcodes, AMAprinciples}. Ensuring consistent adherence to these principles across diverse tasks and settings remains a complex and unresolved problem. Moreover, the dynamic nature of medical practice exacerbates alignment difficulties, as LLMs must remain up-to-date with evolving standards, treatments, and ethical considerations.

Another great challenge in aligning medical LLMs stems from the lack of robust tools for interpreting or explaining their behavior. Current interpretability methods often fail to provide faithful or comprehensive insights into how LLMs process inputs or generate outputs, leaving critical gaps in understanding their reasoning pathways. This opacity poses serious risks in medical applications, where harmful outputs—such as misinformation or unethical recommendations—require precise diagnosis of their root causes to enable targeted alignment interventions.
For example, LLMs may generate plausible but incorrect medical advice due to unintended associations formed during training, yet without reliable interpretability tools, it becomes challenging to identify whether such behavior results from training data biases, spurious correlations, or architectural flaws. 

In addition to the challenges of use and fine-tuning, evaluating the alignment of LLMs is a critical aspect of ensuring their safety and reliability, particularly in the medical domain. Evaluation methodologies face several confounding factors, such as prompt-sensitivity \cite{sclar2023quantifying, mizrahi2024state}, test-set contamination \cite{tirumala2022memorization}, and biases in human-based assessment frameworks \cite{pandey2022modeling}. For example, the performance of an LLM can vary significantly based on the phrasing of the prompt, complicating the consistent measurement of alignment quality. Furthermore, targeted training and fine-tuning of LLMs can lead to overfitting on evaluation benchmarks, giving an inflated impression of alignment without addressing underlying issues. 



Beyond technical and normative alignment, AI-clinician interaction and alignment is also a fundamental aspect of implementing AI and LLM solutions in medical practice. A symbiotic collaboration between AI and clinicians, achieved through consensus finding or disagreements cross-validation, would significantly improve the quality of patient care and reduce clinical operation costs \cite{sanchez2023ai}. Several challenges need to be addressed to ensure effective and seamless integration of AI and LLMs into clinical workflows. 
Complex and non-intuitive interfaces can impede effective interaction between clinicians and AI tools, adversely affecting adoption and user experience. 
Skepticism about the reliability of AI predictions may arise among clinicians, particularly when models lack transparency, interpretability, or validation in real-world settings \cite{ostrowska2024trust}. 
Incorporating AI tools into existing clinical workflows can be challenging, potentially causing disruptions and diminishing efficiency. 
Issues concerning patient privacy, data security, and accountability in cases of erroneous AI predictions can affect clinicians' willingness to employ these tools. 
The performance of AI tools can be compromised by using inaccurate, incomplete, or outdated data, reducing their utility and reliability in clinical settings. 
A gap in AI understanding among clinicians and insufficient medical knowledge among AI developers can lead to less-than-optimal AI tool development and implementation. 
Inadequate training and education regarding AI tools can hinder clinicians' ability to utilize and interpret outputs from them effectively. 
The integration of AI tools into clinical practice might alter the dynamics of patient-clinician relationships, potentially impacting trust, communication, and the overall quality of care. 
Advanced AI models, such as deep neural networks, often lack transparency, making it challenging to understand and anticipate their behaviors and underlying decision-making processes.

\section*{Building Public Confidence in the Safety of AI for Clinical Practice, Healthcare System Operation and Social Impact}\label{section3}
In the previous sections, we highlighted that adoption of AI and LLMs  in real-world medical settings remains limited due to persistent safety concerns. In this section, we explore potential strategies to foster trust and build confidence among patients, clinicians, and the public in the safety of AI within clinical practice, healthcare operations, and its broader social impact.

\subsection*{Confidence in AI safety for clinical practice}

In clinical practice, AI safety concerns exist in various forms, each carrying distinct risks. Although AI and LLMs are usually designed and functioning as assistive tools to support clinical decisions (e.g., diagnostic-AI, clinical monitoring, etc.\cite{tu2024towards, chen2023framework, barnett2023real}),  unrobust, unsafe, and unexplainable black-box AI models could potentially result in \textit{disease misdiagnosis and subsequent mismanagement} \cite{lee2019explainable}. 
It’s been known that AI models with high predictive performance can be \textit{poorly calibrated}, and calibration-reporting guidelines, despite well adopted in the medical community \cite{collins2015transparent}, are rarely followed in the engineering research studies. 
\textit{Uncertainty measures} of predicted outcomes, parameters, and statistics (e.g., confidence intervals) are also commonly under-reported in incomplete and/or inconsistent ways \cite{kompa2021second}. 
\textit{Poor generalization} in AI systems can lead to misdiagnoses and inappropriate treatments, impacting patient health resource allocation and eroding trust in AI technologies. 
Biased models exacerbate these issues, particularly affecting underrepresented groups, leading to misdiagnoses, inappropriate treatments, and widening health disparities. 
Lastly, the challenge of fine-tuning of AI models using local data is essential to ensure equitable healthcare outcomes, underscoring the need for models sensitive to \textit{diverse patient populations and healthcare settings}. 
Each of these risks highlights the critical need for rigorous testing, validation, and ethical considerations in the deployment of AI in medicine in order to build confidence among the users\cite{yu2018artificial, li2022domain, ricci2023towards, 9735278}.

\subsection*{Confidence in AI safety for healthcare system operations}
The integration of AI and LLMs in healthcare systems, such as hospital operations, community clinic operations, pharmacy operations and health insurance, brings various safety issues that could lead to risks and dangers. 
\textit{Data security and privacy} is one of the largest issues when adopting AI and LLMs in healthcare operations, as AI systems require access to vast amounts of sensitive patient data. There is a risk of data breaches, unauthorized access, or misuse of data, which can lead to privacy violations and compromise patient confidentiality \cite{murdoch2021privacy, thapa2021precision}. Therefore, significant improvement of data protection will be vital to build confidence and trust in using AI and LLMs in daily healthcare system operation. 
Another issue is \textit{robustness}, which refers to the possibility that AI systems might make errors in diagnosis or treatment recommendations due to limitations in their algorithms or training data. These errors can have serious consequences for patient health \cite{balagurunathan2021requirements}. 
Ensuring that AI systems comply with existing healthcare \textit{regulations and standards} is crucial. Non-compliance can lead to legal and ethical issues.
Additionally, \textit{interoperability and system integration} pose challenges in healthcare as integrating AI and LLMs into existing systems can be complicated due to inter-operator discrepancies and differences in data standards, which can lead to inefficiencies and errors in patient care \cite{liu2019high}.
\textit{Resource allocation} should also be taken into consideration, as AI-driven decisions might influence the distribution of resources in healthcare, potentially prioritizing certain groups or treatments over others, which could exacerbate inequalities \cite{formosa2022medical}.
Apart from these technical problems, \textit{dependence and skill degradation} among healthcare workers can occur due to over-reliance on AI as well as LLMs, leading to a decrease in their decision-making skills \cite{salla2018ai}.

\subsection*{Confidence in social impacts}
AI safety issues in medicine when presented on a societal scale, can lead to a range of broader risks and challenges. Here we outline general issues that may arise from policy level down to population level in Table \ref{table33}. These include public trust and perception, the ethical and societal norms, and the bias, fairness issues and more. To foster public confidence in the application of AI in medicine, it is essential to address these social issues with careful consideration and meticulous attention to detail.

\begin{table*}[htbp]
\caption{Social risks to consider when building confidence in AI for medicine}\label{table33}
    \centering
    \scalebox{0.92}{
        \begin{tabular}{p{3.5cm}|p{13.5cm}}
            \hline 
            \hline 
            {Widening Health Disparities} & If AI technologies are primarily accessible to well-resourced healthcare systems or populations, it could exacerbate existing health disparities. This has been observed in other technological advances and is termed the Matthew principle. Those without access to AI-enabled healthcare might receive lower quality care, widening the health outcome gap between different socio-economic groups \cite{currie2022social}.\\\hline	

            {Economic Impacts}& The costs associated with developing, maintaining, and updating AI systems in healthcare could be substantial. If these costs are not managed effectively, they could strain healthcare budgets, potentially diverting funds from other critical healthcare services.\\\hline	

            {Workforce Displacement and Transformation}& The integration of AI in healthcare might lead to job displacement or require significant transformation of healthcare roles. This could result in job loss or require substantial retraining for healthcare professionals, impacting employment and the structure of the healthcare workforce.\\\hline	

            {Policy and Regulatory Challenges}& As AI continues to advance, it may outpace current regulations and ethical guidelines, leading to legal and ethical dilemmas. Ensuring that policies keep up with technological advancements is crucial but challenging, and failure to do so could lead to societal risks \cite{pesapane2021legal}.\\\hline	
            
            {Influence on Healthcare Policies and Priorities}& AI's recommendations and predictions could unduly influence healthcare policies and funding priorities, potentially favoring certain diseases, treatments, or research areas over others, based on algorithmic outputs rather than comprehensive public health needs.\\\hline	
            
            {Misinformation and Manipulation Risks}& There's a risk of AI being used to spread misinformation about health issues or being manipulated to provide misleading information, which can have widespread negative effects on public health and safety \cite{federspiel2023threats}.\\\hline	
            
            {Global Inequities}& The uneven distribution and adoption of AI in healthcare across different countries can exacerbate global health inequities. Countries with limited access to AI technology may fall behind in healthcare advancements \cite{celi2022sources}.\\\hline	
            
            {Dependency and System Vulnerabilities}& Over-reliance on AI systems can make healthcare infrastructure vulnerable to technical failures, cyberattacks, or other disruptions, potentially leading to widespread healthcare service interruptions \cite{quinn2021trust}.
            \\\hline	
    \end{tabular}}
\end{table*}

\section*{CONCLUSION}


In this review, we discussed the potential safety issues of AI in healthcare, highlighting LLMs' rapid advancements alongside the significant safety challenges that impede its broader adoption. We identified key concerns related to AI reliability, including data harmonization, model calibration, generalization, biases, and adaption in new population, as well as alignment issues, such as ensuring AI adheres to human-defined objectives. Furthermore, we delved into the specific risks posed by LLMs in medical applications, such as hallucinations, privacy concerns, and difficulties in processing complex logic. Ultimately, while AI holds transformative potential for healthcare, addressing these safety risks and establish trust among the general public is crucial for ensuring the ethical, reliable, and secure deployment of AI technologies in real-world medical settings.



\newpage

\newpage


\section*{RESOURCE AVAILABILITY}


\subsection*{Lead contact}


Requests for further information and resources should be directed to and will be fulfilled by the lead contact, Dianbo Liu (dianbo@nus.edu.sg).

\subsection*{Materials availability}


This study did not generate new materials.



\begin{itemize}
    \item  All the data and papers reported in this review will be shared by the lead contact upon request.
    \item This paper does not report original code.
    \item Any additional information required to reanalyze the papers reported in this paper is available from the lead contact upon request.    
\end{itemize}

\section*{ACKNOWLEDGMENTS}


This study was funded by  National University of Singapore. The funder played no role in study design, data collection, analysis and interpretation of data, or the writing of this manuscript.

\section*{AUTHOR CONTRIBUTIONS}


\begin{itemize}
    \item {Xiaoye Wang}: Investigation; Visualization; Writing – original draft; Writing – review \& editing (Equal Contribution)
    \item {Nicole Xi Zhang}: Investigation; Visualization; Writing – original draft; Writing – review \& editing (Equal Contribution)

    \item {Hongyu He}: Investigation; Writing – original draft; Writing – review \& editing

    \item {Trang Nguyen}: Investigation; Writing – review \& editing

    \item {Kun-Hsing Yu}: Medical inputs; Writing – review \& editing

    \item {Hao Deng}: Medical inputs; Writing – review \& editing

    \item {Cynthia Brandt}: Medical inputs; Writing – review \& editing

    \item {Danielle S. Bitterman}: Medical inputs; Writing – review \& editing

    \item {Ling Pan}: Machine learning inputs; Writing – review \& editing

    \item {Ching-Yu Cheng}: Medical inputs; Writing – review \& editing

    \item {James Zou}: Medical inputs; Writing – review \ editing

    \item {Dianbo Liu}: Conceptualization; Investigation; Visualization; Supervision; Project administration; Writing – original draft; Writing – original draft
\end{itemize}

\section*{DECLARATION OF INTERESTS}


All authors declare no financial or non-financial competing interests.

\section*{DECLARATION OF GENERATIVE AI AND AI-ASSISTED TECHNOLOGIES}



All the scientific contents in this work including but are not limited to listed facts, explanation, views, ideas and citations are written by authors of this article. During the preparation of this work the ChatGPT-4o was used in order to polish and refine languages especially the grammars. After using this tool, the authors reviewed and edited the content as needed and take full responsibility for the content of the publication.

\newpage


\bibliography{references}

\bigskip


\end{document}